\documentclass[12pt]{iopart}

\usepackage{graphicx}
\usepackage{amsfonts}

\begin{document}

\title{Suppressing traffic-driven epidemic spreading by efficient routing protocol}

\author{Han-Xin Yang$^{1}$ and Zhi-Xi Wu$^{2}$}
\address{$^{1}$Department of Physics, Fuzhou University, Fuzhou
350108, China}
\address{$^{2}$Institute of Computational Physics and Complex Systems, Lanzhou University, Lanzhou, Gansu 730000, China}

\begin{abstract}
Despite extensive work on the interplay between traffic dynamics and
epidemic spreading, the control of epidemic spreading by routing
strategies has not received adequate attention. In this paper, we
study the impact of efficient routing protocol on epidemic
spreading. In the case of infinite node-delivery capacity, where the
traffic is free of congestion, we find that that there exists
optimal values of routing parameter, leading to the maximal epidemic
threshold. This means that epidemic spreading can be effectively
controlled by fine tuning the routing scheme. Moreover, we find that
an increase in the average network connectivity and the emergence of
traffic congestion can suppress the epidemic outbreak.
\end{abstract}


 \maketitle
\tableofcontents
\section{Introduction} \label{sec:intro}

Epidemic spreading~\cite{1,2,3,3.1,4,5,6,7,8,9,10,11} and traffic
dynamics~\cite{12,13,14,15,16,17,18,19} on complex networks have
attracted much attention in the past decade. For a long time, the
two types of dynamical processes have been studied independently.
However, in many cases, epidemic spreading is relied on the process
of transportation. For example, a computer virus can spread over
Internet via data transmission~\cite{air1,air2}. Another example is
that air transport tremendously accelerates the propagation of
infectious diseases among different countries.

The first attempt to incorporate traffic into epidemic spreading is
based on metapopulation model~\cite{m1,m2,m3,m4,m5,m6,m7,m8,m9,m10}.
This framework describes a set of spatially structured interacting
subpopulations as a network, whose links denote the traveling path
of individuals across different subpopulations. Each subpopulation
consists of a large number of individuals. An infected individual
can infect other individuals in the same subpopulation. In a recent
work, Meloni $et$ $al.$ proposed another traffic-driven epidemic
spreading model~\cite{Meloni}, in which each node of a network
represents a router in the Internet and the epidemic can spread
between nodes by the transmission of packets. A susceptible node
will be infected with some probability every time it receives a
packet from an infected neighboring node.

Despite broad interests in traffic-driven epidemic spreading, the
control of epidemic spreading by routing strategies has received
little attention. In another recent work, Meloni $et$ $al$. observed
that when travelers decide to avoid locations with high levels of
prevalence, this self-initiated behavioral change may enhance
disease spreading~\cite{avoid}. Later, Yang $et$ $al$. found that
epidemic spreading can be effectively controlled by a local routing
strategy~\cite{yang}. In the local routing protocol, each node does
not know the whole network's topological information and the packet
is forwarded to a neighboring node $i$ with a probability that is
proportional to the power of $i$'s degree~\cite{wang}. It is noted
that in the local traffic routing, the average traveling time of a
packet $\langle T \rangle$ is proportional to the network size
$N$~\cite{pre}. However, in global routing protocols such as the
shortest-path routing, $\langle T \rangle$ usually increases
approximately logarithmically with $N$~\cite{rmp}. Thus, from the
view of transmission time, global routing protocols may be superior
to local routing protocols.

So far, the control of traffic-driven epidemic spreading by a global
routing protocol has not been studied. To address the above issue,
we consider an efficient routing strategy proposed by Yan $et$ $al.$
~\cite{yan}. In the efficient routing protocol, each node in a
network is assigned a weight that is proportional to the power of
its degree, where the power exponent $\alpha$ is a tunable
parameter. The efficient path between any two nodes is corresponding
to the route that makes the sum of the nodes' weight (along the
path) minimal. It has been proved that the traffic throughput of the
network can be greatly improved by employing the efficient routing
strategy as compared to the shortest-path strategy~\cite{yan}. In
this paper, we intend to study how the the efficient routing
protocol affects traffic-driven epidemic spreading. Our preliminary
results have shown that there exists an optimal value of $\alpha$,
leading to the maximal epidemic threshold.

The paper is organized as follows. In Sec.~\ref{sec:methods}, we
formalize the problem by introducing the efficient routing strategy into
traffic-driven epidemic spreading. In Sec.~\ref{sec:main results},
we investigate the epidemic spreading on scale-free networks by
considering two cases of node-delivering capacity, i.e., infinite
capacity and finite capacity. Finally, conclusions and discussions are presented in Sec.~\ref{sec:discussion}.

\section{Model and Methods} \label{sec:methods}

Following the work of Meloni $et$ $al.$~\cite{Meloni}, we incorporate
the traffic dynamics into the classical susceptible-infected-susceptible
model~\cite{SIS} of epidemic spreading as follows.

(i) {\em Efficient routing protocol.} In a network of size $N$, at
each time step, $\lambda N$ new packets are generated with randomly
chosen sources and destinations (we call $\lambda$ as the
packet-generation rate), and each node can deliver at most $C$
packets towards their destinations. For any path between nodes $i$
and $j$, $P(i \rightarrow j) : = i \equiv x_{1}, x_{2}, \cdots ,
x_{n}\equiv j$, we define

\begin{equation}\label{Eq1}
L\left(P(i \rightarrow j):\alpha\right)=\sum_{l=1}^{n}k(x_{l})^{\alpha},
\end{equation}
where $k(x_{l})$ is the degree of node $x_{l}$ and $\alpha$ is a
tunable parameter. For any given $\alpha$, the efficient path
between $i$ and $j$ is corresponding to the route that makes the sum
$L\left(P(i \rightarrow j):\alpha\right)$ minimum. Packets are delivered
following the efficient path. When $\alpha=0$, the efficient path
recovers the traditional shortest path. Once a packet reaches its
destination, it is removed from the system. The queue length of each
node is assumed to be unlimited and the first-in-first-out principle
holds for the queue.

(ii) {\em Epidemic dynamics.} After a transient time, the total
number of delivered packets at each time will reach a steady value,
then an initial fraction of nodes $\rho_{0}$ is set to be infected
(e.g., we set $\rho_{0}=0.1$ in numerical experiments). The
infection spreads in the network {\em through packet exchanges}.
Each susceptible node has the probability $\beta$ of being infected
every time it receives a packet from an infected neighbor. The
infected nodes recover at rate $\mu$ (we set $\mu=1$ in this paper).

\section{Main results and Analysis} \label{sec:main results}

In the following, we carry out simulations systematically by
employing traffic-driven epidemic spreading on the
Barab\'{a}si-Albert (BA) scale-free networks~\cite{ba}. The size of
the BA network is set to be $N=2000$. In the case where the
node-delivering capacity is infinite ($C\rightarrow \infty$),
traffic congestion will not occur in the network. When the
node-delivering capacity is finite, traffic congestion can occur if
the packet-generating rate $\lambda$ exceeds a critical value~\cite{yang}.
Therefore infinite and finite node-delivering capacity are
considered respectively in the following sections.

\subsection{Infinite node-delivering capacity}

\begin{figure}
\begin{center}
\includegraphics[width=110mm]{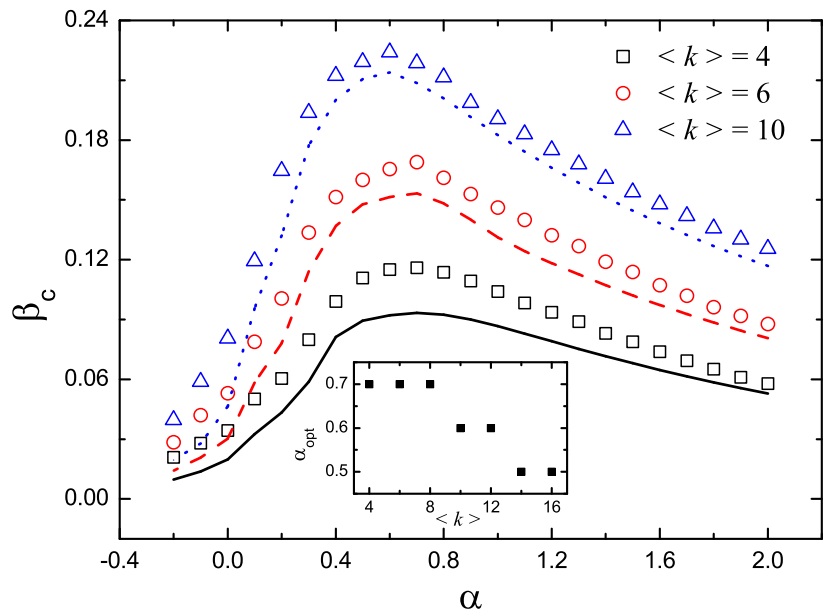}
\caption{ The epidemic threshold $\beta_{c}$ as a function of
$\alpha$ for different values of $\langle k \rangle$. The symbols
are from numerical simulations. The solid, dashed, and dotted curves
correspond to the theoretical prediction from Eq.~(2) for $\langle
k\rangle=$ 4, 6 and 10, respectively. The inset shows the optimal
value $\alpha_{\mathrm{opt}}$ of the routing parameter as a function
of node degree $\langle k \rangle$. The packet-generation rate
$\lambda=1$. Each data point results from an average over 100
different realizations.}\label{fig2}
\end{center}
\end{figure}

Previous studies have shown that there exists an epidemic threshold
$\beta_{c}$, below which the epidemic goes
extinct~\cite{Meloni,yang}. Figure~\ref{fig2} shows the dependence
of $\beta_{c}$ on $\alpha$ for different values of the average
degree $\langle k\rangle$ of the network. We find that for each
value of $\langle k \rangle$, there exists an optimal value of
$\alpha$, hereafter denoted by $\alpha_{\mathrm{opt}}$, leading to
the maximum $\beta_{c}$. The inset of Fig.~\ref{fig2} shows that
$\alpha_{\mathrm{opt}}$ decreases from 0.7 to 0.5 as $\langle k
\rangle$ increases from 4 to 16.

According to the analysis in Ref.~\cite{Meloni}, the epidemic
threshold for uncorrelated networks is
\begin{equation}
\beta_{c}=\frac{\langle b_{\mathrm{alg}} \rangle}{\langle
b_{\mathrm{alg}}^{2} \rangle}\frac{1}{\lambda N},
\end{equation}
where $b_{\mathrm{alg}}$ is the efficient algorithmic betweenness of
a node~\cite{alg1,alg2} and $\langle \cdot \rangle$ denotes the
average of all nodes. The efficient algorithmic betweenness of a
node represents the average number of packets passing through that
node at each time step when the packet-generation rate
$\lambda=1/N$. In this paper, the efficient algorithmic be-
tweenness of a node $k$ can be calculated as
\begin{equation}
b_{\mathrm{alg}}^{k}=\frac{1}{N(N-1)}\sum_{i\neq
j}\frac{\sigma_{ij}(k)}{\sigma_{ij}},
\end{equation}
where $\sigma_{ij}$ is the total number of efficient paths going
from $i$ to $j$, and $\sigma_{ij}(k)$ is the number of efficient
paths going from $i$ to $j$ and passing through $k$. From
Fig.~\ref{fig2}, one can see that the theoretical predictions agree
with numerical results qualitatively.

\begin{figure}
\begin{center}
\includegraphics[width=92mm]{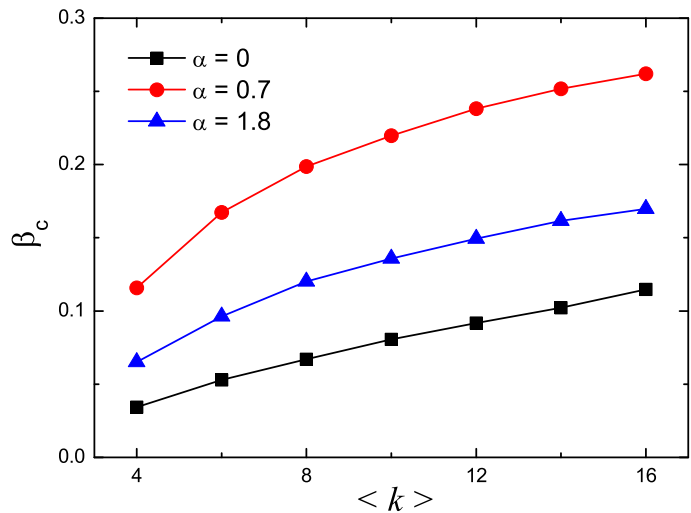}
\caption{ The epidemic threshold $\beta_{c}$ as a function of
$\langle k \rangle$ for different values of $\alpha$. The
packet-generation rate $\lambda=1$. Each data point results from an
average over 100 different realizations.}\label{fig3}
\end{center}
\end{figure}

Next, we study the effect of the average degree of the network on
the traffic-driven epidemic spreading. Figure~\ref{fig3} shows the
epidemic threshold $\beta_{c}$ as a function of the average degree
of the network $\langle k \rangle$ for different values of $\alpha$.
From Fig.~\ref{fig3}, we find that for each value of $\alpha$,
$\beta_{c}$ increases with $\langle k \rangle$, in contrast to the
behavior of spreading dynamics in the absence of
traffic~\cite{absent}. This phenomenon can be understood as follows.
An increase in the average degree of the network shortens the
average time steps that a packet spends traveling from its source to
its destination and decreases the number of packages passing through
each node, leading to a decrease in the infection probability of
each node.

\begin{figure}
\begin{center}
\includegraphics[width=96mm]{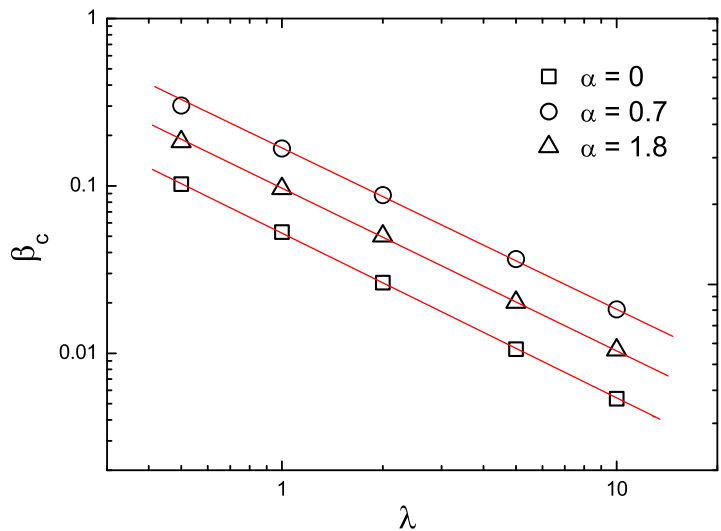}
\caption{The epidemic threshold $\beta_{c}$ as a function of the
packet-generation rate $\lambda$ for different values of $\alpha$.
The average degree of the network $\langle k \rangle=6$. The slope
of the fitted line is about -1. Each data point results from an
average over 100 different realizations. }\label{fig4}
\end{center}
\end{figure}

Figure~\ref{fig4} shows the epidemic threshold $\beta_{c}$ as a
function of the packet-generation rate $\lambda$ for different
values of $\alpha$. One can see that $\beta_{c}$ scales inversely
with $\lambda$, as predicted by Eq. (2), indicating that the
increase of traffic flow facilitates the outbreak of epidemic. The
similar result has also been found in Ref.~\cite{Meloni}.

\begin{figure}
\begin{center}
\includegraphics[width=96mm]{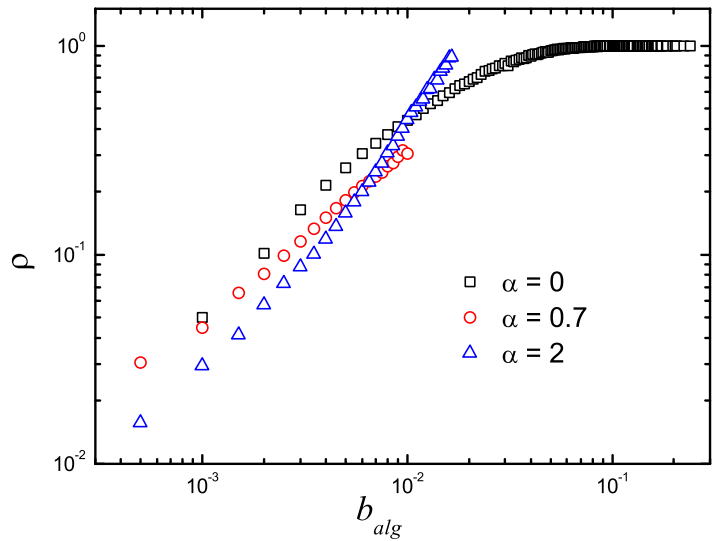}
\caption{The infection probability of nodes $\rho$ as a function of
 the algorithmic betweenness $b_{\mathrm{alg}}$ for different values of $\alpha$. The average
degree of the network $\langle k \rangle=6$. Each data point results
from an average over 100 different realizations. }\label{fig6}
\end{center}
\end{figure}

An interesting issue is how the algorithmic betweenness
$b_{\mathrm{alg}}$ affects the infection probability of nodes
$\rho$. Figure~\ref{fig6} features the dependence of $\rho$ on
$b_{\mathrm{alg}}$ for different values of $\alpha$. From
Fig.~\ref{fig6}, one can observe that the algorithmic betweenness of
the nodes is positively correlated with the risk of them being
infected.

\subsection{Finite node-delivering capacity}

\begin{figure}
\begin{center}
\includegraphics[width=100mm]{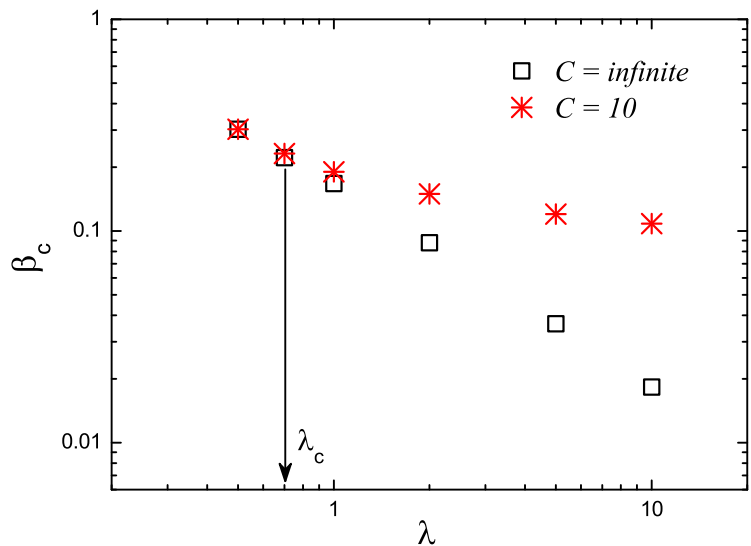}
\caption{ The epidemic threshold $\beta_{c}$ as a function of the
packet-generation rate $\lambda$ for finite and infinite $C$. The
average degree of the network $\langle k \rangle=6$ and the routing
parameter $\alpha=0.7$. The critical packet-generating rate for
$C=10$ is $\lambda_{c}\approx0.7$. Each data point results from an
average over 100 different realizations. }\label{fig7}
\end{center}
\end{figure}

In the case of finite node-delivering capacity, traffic congestion
occurs when the packet-generating rate exceeds a critical value
$\lambda_{c}$, which can can be estimated as~\cite{alg2,zhao},
\begin{equation}
\lambda_{c}=\frac{C}{N b_{\mathrm{alg}}^{\mathrm{max}}},
\end{equation}
where $b_{\mathrm{alg}}^{\mathrm{max}}$ is the largest algorithmic betweenness of the network.

Figure~\ref{fig7} shows the epidemic threshold $\beta_{c}$ as a
function of the packet-generation rate $\lambda$ for finite and
infinite $C$. One can see that when $\lambda \leq \lambda_{c}$,
$\beta_{c}$ is identical for both cases of the finite and infinite
delivery capacities. However, for $\lambda>\lambda_{c}$, $\beta_{c}$
is larger in the case of finite capacity than that in the infinite
capacity case, indicating that traffic congestion can suppress the
outbreak of epidemic. This result is consistent with that in
Ref.~\cite{Meloni}. The above phenomenon can be explained as
follows. Once a node becomes congested, it cannot deliver the total
packets in its queue at each time step. A decrease in the number of
delivered packets can help nodes reduce the probability of being
infected.

\begin{figure}
\begin{center}
\includegraphics[width=100mm]{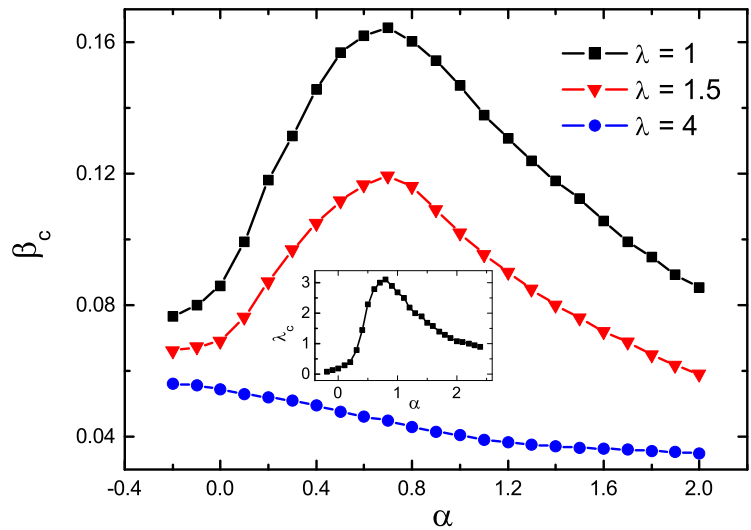}
\caption{The epidemic threshold $\beta_{c}$ as a function of
$\alpha$ for different values of the packet-generation rate
$\lambda$. The average degree of the network $\langle k \rangle=6$
and the node-delivering capacity $C=50$. The inset shows that the
critical packet-generating rate $\lambda_{c}$ as a function of
$\alpha$. When $\lambda=1$, $\lambda>\lambda_{c}$ for $\alpha<0.4$
and $\alpha>2.2$. When $\lambda=1.5$, $\lambda>\lambda_{c}$ for
$\alpha<0.5$ and $\alpha>1.6$. When $\lambda=4$,
$\lambda>\lambda_{c}$ for all the values of $\alpha$. Each data
point results from an average over 100 different realizations.
}\label{fig8}
\end{center}
\end{figure}

Figure~\ref{fig8} shows the epidemic threshold $\beta_{c}$ as a
function of $\alpha$ for different values of the packet-generation
rate $\lambda$. One can observe that for small values of $\lambda$
(e.g., $\lambda=1$ or $\lambda=1.5$), there exists an optimal value
of $\alpha$, leading to the maximal $\beta_{c}$. However, when
$\lambda$ is large enough (e.g., $\lambda=4$), $\beta_{c}$ decreases
with the increase of $\alpha$. The inset of Fig.~\ref{fig8} shows
the critical packet-generating rate $\lambda_{c}$ as a function of
$\alpha$. One can see $\lambda_{c}$ is maximized by an optimal value
of $\alpha$.

\section{Conclusions and Discussions} \label{sec:discussion}

In conclusion, we have studied the impact of efficient routing
protocol on traffic-driven epidemic spreading. We find that the
epidemic threshold increases with the average degree of the network
when other parameters are fixed. Besides, we find that nodes with
larger algorithmic betweenness are more likely to be infected. Both
analytic and numerical results show that, there exists an optimal
value of routing parameter, leading to the maximal epidemic
threshold. This means that epidemic spreading can be controlled by
fine tuning the routing scheme. We hope our results can be useful to
understand and control spreading dynamics.

\section*{Acknowledge}
This work was supported by the National Natural Science Foundation
of China (Grants No. 11247266, No. 11005051, and No. 11135001), the
Natural Science Foundation of Fujian Province of China (Grant No.
2013J05007), and the Research Foundation of Fuzhou University (Grant
No. 0110-600607).


\section*{References}
\begin{thebibliography}{100}

\bibitem{1} Pastor-Satorras R and Vespignani A 2001 Phys. Rev. Lett.
\textbf{86} 3200

\bibitem{2}
Newman M E J 2002 Phys. Rev. E \textbf{66} 016128

\bibitem{3}
Barth\'{e}lemy M, Barrat A, Pastor-Satorras R and Vespignani A 2004
Phys. Rev. Lett. \textbf{92} 178701

\bibitem{3.1} Gross T, Dommar D'Lima C J and Blasius B 2006 Phys. Rev. Lett. \textbf{96} 208701

\bibitem{4}
Yan G, Fu Z Q, Ren J and Wang W X 2007 Phys. Rev. E \textbf{75}
016108

\bibitem{5} Kitsak M, Gallos L K, Havlin S, Lijeros F, Muchnik L, Stanley H E
and Makse H A 2010  Nat. Phys. \textbf{6} 888

\bibitem{6}
Parshani R, Carmi S and Havlin S, Phys. Rev. Lett. 2010 \textbf{104}
258701

\bibitem{7} Castellano C and Pastor-Satorras R 2010
Phys. Rev. Lett. \textbf{105} 218701

\bibitem{8} Karrer B and Newman M E J 2011 Phys. Rev. E \textbf{84} 036106
\bibitem{9} Castellano C and Pastor-Satorras R 2012 Sci. Rep. \textbf{2} 372
\bibitem{10} Ruan Z, Tang M and Liu Z 2012 Phys. Rev. E \textbf{86} 036117
\bibitem{11} Bogu\~{n}\'{a} M, Castellano C and Pastor-Satorras R 2013 Phys. Rev. Lett. \textbf{111} 068701


\bibitem{12} Echenique P, G\'{o}mez-Garde\~{n}es J and Moreno Y 2004 Phys. Rev. E \textbf{70} 056105
\bibitem{13} Echenique P, G\'{o}mez-Garde\~{n}es J and Moreno Y 2005 Europhys. Lett. \textbf{71} 325
\bibitem{14} Meloni S, G\'{o}mez-Garde\~{n}es J, Latora V and Moreno 2008 Phys. Rev. Lett. \textbf{100} 208701
\bibitem{15} Bogu\~{n}\'{a} M, Krioukov D and Claffy K C 2009 Nat. Phys. \textbf{5} 74
\bibitem{16} Tang M, Liu Z, Liang X and Hui P M 2009 Phys. Rev. E \textbf{80} 026114
\bibitem{17} Meloni S and G\'{o}mez-Garde\~{n}es J 2010 Phys. Rev. E \textbf{82} 056105
\bibitem{18} Yang H X, Wang W X, Xie Y B, Lai Y C and Wang B H 2011 Phys. Rev. E \textbf{83} 016102
\bibitem{19} Morris R G and Barthelemy M 2012 Phys. Rev. Lett. \textbf{109} 128703


\bibitem{air1} Viboud C, Bj{\o}rnstad O N, Smith D L, Simonsen L, Miller M
A and Grenfell B T 2006 Science \textbf{312} 447

\bibitem{air2} Tizzoni M, Bajardi P, Poletto C, Ramasco J J, Balcan D, Gon\c{c}alves B, Perra N, Colizza V and
Vespignani A 2012 BMC Med. \textbf{10} 165





\bibitem{m1} Colizza V, Barrat A, Barth\'{e}lemy M and Vespignani A 2006 Proc.
Natl Acad. Sci. USA \textbf{103} 2015
\bibitem{m2} Colizza V, Pastor-Satorras R and Vespignani A 2007 Nat. Phys. \textbf{3}
276
\bibitem{m3} Colizza V and Vespignani A 2007 Phys. Rev. Lett. \textbf{99} 148701
\bibitem{m4} Colizza V and Vespignani A 2008 J. Theor. Biol. \textbf{251} 450
\bibitem{m5} Gautreau A, Barrat A and Barth\'{e}lemy M 2008 J. Theor. Biol. \textbf{251} 509
\bibitem{m6} Balcan D, Colizza V, Gon\c{c}alves B, Hu H, Ramasco J J and
Vespignani A 2009 Proc. Natl Acad. Sci. USA \textbf{106} 21484
\bibitem{m7} Xuan Q, Du F, Yu L and Chen G 2013 Phys. Rev. E \textbf{87} 032809
\bibitem{m8} Balcan D and Vespignani A 2011 Nat. Phys. \textbf{7} 581
\bibitem{m9} Ruan Z, Hui P, Lin H and Liu Z 2013 Eur. Phys. J. B \textbf{86} 13
\bibitem{m10} Liu S Y, Baronchelli A and Perra N 2013 Phys. Rev. E \textbf{87} 032805

\bibitem{Meloni} Meloni S, Arena A and Moreno Y 2009 Proc. Natl Acad. Sci. USA
\textbf{106} 16897




\bibitem{avoid} Meloni S, Perra N, Arenas A, G\'{o}mez S, Moreno Y and
Vespignani A 2011 Sci. Rep. \textbf{1} 62
\bibitem{yang} Yang H X, Wang W X, Lai Y C, Xie Y B and Wang B H 2011 Phys.
Rev. E \textbf{84} 045101(R)
\bibitem{wang} Wang W X, Wang B H, Yin C Y, Xie Y B and Zhou T 2006 Phys. Rev. E \textbf{73} 026111
\bibitem{pre} Fronczak A and Fronczak P 2009 Phys. Rev. E \textbf{80} 016107
\bibitem{rmp} Albert R and Barabasi A L 2002 Rev. Mod. Phys. \textbf{74} 47
\bibitem{yan} Yan G, Zhou T, Hu B, Fu Z Q and Wang B H 2006 Phys.
Rev. E \textbf{73} 046108

\bibitem{SIS} Bailey N T J 1975 \textit{ The Mathematical Theory of Infectious
Diseases} (Griffin, London)

\bibitem{ba} Barabasi A L and Albert R 1999 Science \textbf{286} 509

\bibitem{alg1} Arenas A, D\'{\i}az-Guilera A and Guimer\`{a} R 2001 Phys. Rev.
Lett. \textbf{86} 3196

\bibitem{alg2} Guimer\`{a} R, D\'{\i}az-Guilera A, Vega-Redondo F, Cabrales A
and Arenas A 2002 Phys. Rev. Lett. \textbf{89} 248701

\bibitem{absent} Pastor-Satorras R and Vespignani A 2002 Phys. Rev. E \textbf{65} 035108(R)

\bibitem{zhao} Zhao L, Lai Y C, Park K and Ye N 2005 Phys. Rev. E \textbf{71} 026125




\end{thebibliography}
\end{document}